\documentclass[10pt,preprint]{aastex}

\usepackage{graphicx,apjfonts,emulateapj5,onecolfloat5}
\usepackage{subfigure}
\usepackage{ulem}

\shorttitle{Testing the coincidence problem of dark energy}
\shortauthors{Chen, Y. et al.}

\begin{document}
\twocolumn[


\title{Using a phenomenological model to test the coincidence problem of
       dark energy}


\author{Yun Chen\altaffilmark{1},
        Zong-Hong Zhu\altaffilmark{1},
        J. S. Alcaniz\altaffilmark{2},
        Yungui Gong\altaffilmark{3}
        }


\begin{abstract}
%%%%%%%%%%%%%%%%%%%%%%%%%%%%%%%%%%%%%%%%%%%%%%%%%%%%%%%%%%%%%%%%%%%
By assuming a phenomenological form for the ratio of the dark energy
and matter densities $\rho_X\propto \rho_{m} a^{\xi}$ (Dalal et al.
2001), we discuss the cosmic coincidence problem in light of current
observational data. $\xi$ is a key parameter to denote the severity
of the coincidence problem. In this scenario, $\xi=3$ and $\xi=0$
correspond to $\Lambda$CDM and the self-similar solution without
coincidence problem, respectively. Hence, any solution with a
scaling parameter $0<\xi<3$ makes the coincidence problem less
severe. In addition, the standard cosmology without interaction
between dark energy and dark matter is characterized by
$\xi+3\omega_X=0$, where $\omega_X$ is the equation of state of the
dark energy component, whereas the inequality $\xi+3\omega_X\neq 0$
represents non-standard cosmology.
 We place observational constraints on the parameters
$(\Omega_{X,0},\omega_{X},\xi)$ of this model, where $\Omega_{X,0}$
is the present value of density parameter of dark energy $\Omega_X$,
by using the Constitution Set (397 Supernovae of type Ia data,
hereafter SNeIa), the CMB shift parameter from the 5-year WMAP and
the SDSS baryon acoustic peak. Combining the three samples, we get
$\Omega_{X,0}=0.72\pm0.02$, $\omega_{X}=-0.98\pm0.07$ and
$\xi=3.06\pm0.35$ at $68.3\%$ confidence level. The result shows
that the $\Lambda$CDM model still remains a good fit to the recent
observational data, as well as, the coincidence problem indeed
exists and is quite severe, in the framework of this simple
phenomenological model. We further constrain the model with the
transition redshift (deceleration/acceleration). It shows that if
the transition from deceleration to acceleration happens at the
redshit $z > 0.73$, within the framework of this model, we can
conclude that the interaction between dark energy and dark matter is
necessary.

\end{abstract}


]

\altaffiltext{1}{Department of Astronomy,
                   Beijing Normal University, Beijing 100875, China}

 \altaffiltext{2}{Departmento de Astronomia,
              Observat\'orio Nacional, 20921-400, Rio de Janeiro -- RJ, Brasil}

 \altaffiltext{3}{College of Mathematics and Physics, Chongqing
           University of Posts and Telecommunications, Chongqing 400065, China}


\section{INTRODUCTION}
%%%%%%%%%%%%%%%%%%%%%%%%%%%%%%%%%%%%%%%%%%%%%%%%%%%%%%%%%%%%%%%%%%%
\label{intro} One of the most important issues of modern cosmology
concerns the accelerating expansion of the universe, which has been
discovered and verified by the type Ia supernovae (SNeIa) (Riess et
al. 1998; Perlmutter et al. 1999; Hicken et al. 2009), cosmic
microwave background (CMB) (Spergel et al. 2003) and baryon acoustic
oscillation (BAO) (Eisenstein et al. 2005) observations (also see
recent review: Frieman et al. 2008). After the discovery of this
scenario, a great variety of attempts have been done to explain this
acceleration. The nature of the accelerating expansion is one of the
most outstanding problems of physics and astronomy today. Currently,
the existing mechanisms for cosmic acceleration can be roughly
assorted into three kinds (see the reviews: Copeland et al. 2006;
Caldwell and Kamionkowski 2009): (i) An exotic energy component with
negative pressure, dubbed as dark energy, is introduced in the right
hand side of Einstein equation. The nature of the dark energy is
still unknown. Dark energy models include $\Lambda$CDM model
(Carroll et al. 1992; Riess et al. 1998; Peeples and Ratra 2003),
holographic dark energy model (Li 2004; Wu et al. 2008), chaplygin
gas model (Kamenshchik et al. 2001; Benaoum 2002; Zhu 2004; Zhang
and Zhu 2006) and some scalar field models, such as quintessence
(Caldwell et al. 1998; Zlatev et al. 1999). (ii) The theory of
gravity is modified at the Hubble scale, and the cosmic acceleration
is due to gravity, without the help of any exotic negative pressure
component. Examples of modified gravity theory include braneworld
model (Arkani-Hamed et al. 1998; Antoniadis et al. 1998; Randall and
Sundrum 1999a; Randall and Sundrum 1999b; Dvali et al. 2000;
Deffayet 2001; Zhu and Alcaniz 2005), $f(R)$ gravity (Carroll et al.
2004; Nojiri and Odintsov 2003; Song et al. 2007; Atazadeh et al.
2008; Wu and Zhu 2008) and Cardassian model (Freese and Lewis 2002;
Zhu and Fujimoto 2002, 2003, 2004; Sen and Sen 2003; Mosquera Cuesta
et al. 2008). The above mentioned two mechanisms are based on the
cosmological principle, which claims that our universe is isotropic
and homogeneous at large scales.
 (iii) The local inhomogeneity of our universe is used
to explain the acceleration (see George 2008 and corresponding
references therein).

From the observational point of view, it is well known that flat
models with a very small cosmological term ($\rho_{\Lambda} \lesssim
10^{-47}$ ${\rm{GeV}}^4$) are in good agreement with almost all sets
of cosmological observations, which makes them an excellent
description of the observed Universe. From the theoretical
viewpoint, however, these scenarios are embarrassed by the so-called
cosmological constant problems (Weinberg 1989; Weinberg 2000). One
issue of the cosmological constant problems is to understand in a
natural way why the observed value of the vacuum energy density
$\rho_{\Lambda}$ is so small. As it is known, the theoretical value
of $\rho_{\Lambda}$ is about 120 orders of magnitude larger than the
observed value. This problem is called fine-tuning problem. Another
issue of the cosmological constant problems is the so-called ``why
now?'' or coincidence problem (Zlatev et al. 1999). Briefly put, it
is to understand why $\rho_{\Lambda}$ is not only small, but also
the same order of magnitude as the present mass density of universe.
The present epoch then is the very special time in the history of
the universe, the only period when $\Omega_{\Lambda}\sim\Omega_{m}$.

Although there is no convincing fundamental theory available to
understand why the vacuum energy dominance happened only recently,
several possible approaches have be adopted to explain or alleviate
the coincidence problem. One approach invokes some sorts of
anthropic principle to explain the coincidence problem (Weinberg
2000; Vilenkin 2001; Garriga et al. 2000; Garriga and Vilenkin
2001). The basic idea of the anthropic principle is that there is an
ensemble of universes with different values of the vacuum energy,
most of which do not allow life to develop. Therefore the cosmic
coincidence is explained as follows: the existence of intelligent
life selects only those values of vacuum energy density close to the
observed one. Another possible approach to alleviate the coincidence
problem is to assume that the dark energy is not a cosmological
constant, but is a slow evolving and spatially homogeneous scalar
field, the so-called ``tracker field'' (see the review: Copeland et
al. 2006). However, in such models the resolution of the cosmic
coincidence problem typically leads to a fine-tuning of model
parameter. Recently, an interesting proposal is that the interaction
between the dark energy and dark matter could perhaps alleviate the
coincidence problem. So far, various interaction models have been
put forward and studied (Amendola 2000; Caldera-Cabral 2009), but
the format of interaction ``Q'' still can not be determined from
fundamental physics.

As it is known, in the $\Lambda$CDM scenario, the densities scale as
$\rho_X \propto \rho_m a^3$, while in a theory with no coincidence
problem, one expects $\rho_X \propto \rho_m$. In this paper, we
adopt a simple model, assuming a phenomenological form for the ratio
of the dark energy and matter densities $\rho_X\propto \rho_{m}
a^{\xi}$ (Dalal et al. 2001), to test the cosmic coincidence problem
, wherein $\xi$ denotes the severity of the coincidence problem. In
this phenomenological model, the standard cosmology without
interaction between dark energy and dark matter is characterized by
$\xi+3\omega_X=0$, thereby, $\xi+3\omega_X\neq 0$ denotes
non-standard cosmology. The special cases $\xi=3$ and $\xi=0$
correspond to $\Lambda$CDM and the self-similar solution without
coincidence problem respectively. Hence, any solution with a scaling
parameter $0<\xi<3$ makes the coincidence problem less severe (Pavon
et al. 2004). We place constraints on the parameters
$(\Omega_{X,0},\omega_{X},\xi)$ of this model by using the
Constitution Set (397 SNeIa data), the CMB shift parameter from the
5-year WMAP and the SDSS baryon acoustic peak,where $\Omega_{X,0}$
is the present value of density parameter of dark energy component.
When we work out the value of $\xi$, one can see how severe the
coincidence problem is. In addition, we use the transition redshift
$z_T$ (deceleration/acceleration) to test whether the interaction
between dark energy and dark matter is necessary.

The paper is organized as follows. In Section 2, the basic equations
of the simple phenomenological model are given. Constraints from the
recent observations are illustrated in Section 3. In Section 4,
theoretical constraints from the transition redshift $z_{T}$ are
presented. Finally, we summarize our main conclusions in Section 5.

%%%%%%%%%%%%%%%%%%%%%%%%%%%%%%%%%%%%%%%%%%%%%%%%%%%%%%%%%%%%%%%%%%%
%%%%%%%%%%%%%%%%%%%%%%%%%%%%%%%%%%%%%%%%%%%%%%%%%%%%%%%%%%%%%%%%%%%
\section{The basic equations of a simple phenomenological model}
Although many dark energy models are proposed, the nature of the
dark energy is still unknown. In this paper, we adopted an
alternative approach of using the observational data to constrain
the nature of dark energy with minimal underlying theoretical
assumptions. In particular, we assume a phenomenological form for
the ratio between the dark energy and matter energy densities
\cite{Dalal2001},
\begin{equation}
\label{eq:ro} \rho_X\propto\rho_m a^{\xi},
\end{equation}
or $\Omega_X\propto\Omega_m a^{\xi}$, where $\xi$ is a free
parameter which can be constrained from the observations.

Considering a flat FRW universe with $\Omega_X+\Omega_m=1$
throughout, we obtain
\begin{equation}
\label{eq:ox}
\Omega_X=\frac{\Omega_{X,0}a^{\xi}}{1-\Omega_{X,0}(1-a^{\xi})},
\end{equation}
where $\Omega_{X,0}$ is the present value of $\Omega_X$. According
to the energy conservation equation, one obtains
\begin{equation}
\label{eq:energycon}
\frac{d\rho_{tot}}{da}=\frac{3}{a}(1+\omega_X\Omega_X)\rho_{tot}=0,
\end{equation}
where $\rho_{tot}=\rho_m+\rho_X$ is the total density. Setting
\begin{equation}
\label{eq:mX} \rho_X=\kappa \rho_m a^{\xi},
\end{equation}
based on Eq.(\ref{eq:energycon}), we have
\begin{equation}
\label{eq:rm1}
\frac{d\rho_m}{da}+\frac{3}{a}\rho_m=-(\xi+3\omega_X)\frac{\kappa
a^{\xi-1}}{1+\kappa a^{\xi}}\rho_m,
\end{equation}
where $\kappa$ is a constant. Further, one can read
\begin{equation}
\label{eq:rm2} \frac{d\rho_m}{da}+\frac{3}{a}\rho_m=Q,
\end{equation}
and
\begin{equation}
\label{eq:rX} \frac{d\rho_X}{da}+\frac{3}{a}(1+\omega_X)\rho_X=-Q,
\end{equation}
where the interaction term $Q=-(\xi+3\omega_X)\frac{\kappa
a^{\xi-1}}{1+\kappa a^{\xi}}\rho_m$. The phenomenological
interaction term $Q$ is inspired from the interaction between the
dilaton field $\sigma$ and the matter field in the scalar-tensor
theory of gravity \cite{kaloper},
\begin{equation}
\label{bdaction} {\cal L}= \sqrt{-g}
\left[-\frac{1}{2\kappa^2}{\mathcal R}
-\frac{1}{2}g^{\mu\nu}\partial_\mu \sigma \partial_\nu \sigma
-\zeta(\sigma)^{-2}{\cal L}_{m}(\psi,
\zeta(\sigma)^{-1}g_{\mu\nu})\right].
\end{equation}
For a general coupling function $\zeta(\sigma)$, we get the
interaction term $Q=-3\rho_m H [d(\ln\zeta)/d(\ln a)]/2$
\cite{quiros} which has the desired linear relationship between $Q$
and $\rho_m$. Consequently, $\xi+3\omega_X=0$, corresponding to
$Q=0$, denotes the standard cosmology without interaction between
dark energy and dark matter. In contrast, $\xi+3\omega_X\neq 0$,
corresponding to $Q\neq 0$, represents the non-standard cosmology
with interaction between dark energy and dark matter. Furthermore,
$\xi+3\omega_X>0$, corresponding to $Q<0$, indicates that the energy
is transferred from dark matter to dark energy. On the other hand,
$\xi+3\omega_X<0$, corresponding to $Q>0$, denotes that the energy
is transferred from dark energy to dark matter. The solution to Eq.
(\ref{eq:energycon}) is
\begin{equation}
\label{eq:solution}
\frac{\rho_{tot}}{\rho_0}=\exp[\int_{a}^{1}\frac{da}{a}3(1+\omega_X\Omega_X)].
\end{equation}
By assuming $\omega_X=\textrm{const.}$, we find
\begin{equation}
\label{eq:rotot} \rho_{tot}=\rho_0
a^{-3}[1-\Omega_{X,0}(1-a^{\xi})]^{-3\omega_X/\xi}\;,
\end{equation}
so that
\begin{equation}
\label{eq:e2}
E^2=a^{-3}[1-\Omega_{X,0}(1-a^{\xi})]^{-3\omega_X/\xi},
\end{equation}
where $E=H/H_0$ is the dimensionless Hubble parameter.

This model has three free parameters $(\Omega_{X,0},\omega_X,\xi)$,
where $\Omega_{X,0}$ specifies the current density of dark energy,
$\omega_X$ denotes its equation of state, and $\xi$ presents how
strongly $\Omega_X/\Omega_m$ varies with redshift and quantifies the
severity of the coincidence problem. In addition, the special cases
$\xi=3$ and $\xi=0$ correspond to $\Lambda$CDM and the self-similar
solution without coincidence problem respectively. The parameters
$(\Omega_{X,0}, \omega_X, \xi)$ can be constrained from the
cosmological data as we describe below.

\section{Constraints from the recent observations }

\subsection{Constraints from SNeIa }
As it is known, the first direct evidence for cosmic acceleration
came from SNeIa (Riess et al. 1998; Perlmutter et al. 1999), and
they have provided the strongest constraints on the cosmic
equation-of-state and other cosmological parameters (Riess et al.
2004; Riess et al. 2007; Astier et al. 2006; Wood-Vasey et al. 2007;
Davis et al. 2007; Kowalski et al. 2008; Hicken et al. 2009). Since
they are as bright as typical galaxies when they peak, SNeIa can be
observed to large distances, recommending their utility as
standardizable candles for cosmology. At present, they are the most
effective and mature probe of cosmology. The present analysis uses
the recently compiled ``Constitution set'' of 397 SNeIa data
covering a redshift range $0.015\leq z \leq 1.551$ (Hicken et al.
2009).

\begin{figure}[t]
\centering
  \includegraphics[angle=0,width=90mm]{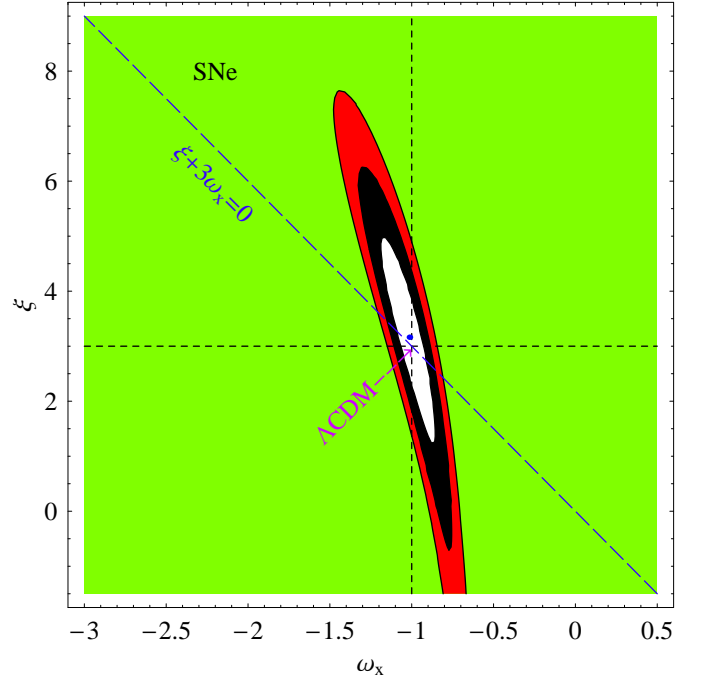}
\caption{
The contours correspond to $68.3\%$, $95.4\%$ and $99.7\%$
confidence levels constrained from SNeIa data analysis in
$(\omega_X, \xi)$ plane. The blue dot marks the best-fit pair
$(\omega_X,\xi)$. The intersection point of the two dashed lines,
namely $(\omega_X, \xi)=(-1, 3)$, corresponds to the $\Lambda$CDM
model. The line $\xi+3\omega_X=0$ corresponds to the standard
cosmology without interaction between dark energy and dark matter.
The best-fit parameters in this case are found to be
$\Omega_{X,0}=0.71\pm0.03$, $\omega_{X}=-1.01\pm0.17$ and
$\xi=3.16\pm1.91$ ($68.3\%$ c.l.).
} \label{fig:SNeCont}
\end{figure}

Constrains from SNeIa can be obtained by fitting the distance
modulus $\mu (z)$. The theoretical distance modulus is ($c = 1$)
\begin{equation}
\label{eq:mu} \mu_{th}(z)=5\log_{10}(D_L(z))+\mu_0,
\end{equation}
where $\mu_0=42.38-5\log_{10}h$. The Hubble-free luminosity distance
is given by
\begin{equation}
\label{eq:DL} D_L(z;\textbf{p},\mu_0)=H_0d_L=(1+z)\int_0^z
\frac{du}{E(u;\textbf{p})},
\end{equation}
where $\textbf{p}\equiv(\Omega_{X,0}, \omega_X, \xi)$ is the
complete set of parameters. The best fit values of parameters can be
determined by minimizing the function
\begin{equation}
\label{eq:x2SN}
\chi^2_{SNe}(\textbf{p},\mu_0)=\sum_{i=1}^{n}\frac{[\mu_{th,i}(\textbf{p},\mu_0;z_i)-\mu_{obs,i}(z_i)]}{\sigma^2_{\mu_i}},
\end{equation}
where $n=397$ is the number of the SNeIa data and $\mu_{obs,i}(z_i)$
is the distance modulus obtained from observations, $\sigma_{\mu_i}$
is the total uncertainty of the SNeIa data. Figure \ref{fig:SNeCont}
shows the probability contours constrained from the 397 SNeIa data
in $(\omega_X , \xi)$ plane. The best-fit parameters in this case
are found to be $\Omega_{X,0}=0.71\pm0.03$,
$\omega_{X}=-1.01\pm0.17$ and $\xi=3.16\pm1.91$ with $68.3\%$
confidence level.

%%%%%%%%%%%%%%%%%%%%%%%%%%%%%%%%%%%%%%%%%%%%%%%%%%%%%%%%%%%%%%%%%%%%%%%%%%%%%%%%%%%%%%%%%%%%%%%%%%%%%%%%%%%%%%%%%%%%%%%%%%%%%%
%%%%%%%%%%%%%%%%%%%%%%%%%%%%%%%%%%%%%%%%%%%%%%%%%%%%%%%%%%%%%%%%%%%%%%%%%%%%%%%%%%%%%%%%%%%%%%%%%%%%%%%%%%%%%%%%%%%%%%%%%%%%%%%%

\subsection{Constraints from BAO }
The recently observed baryon oscillations in the power spectrum of
galaxy correlation function also is a powerful probe to explore dark
energy and constrain cosmological model (Eisenstein et al. 2005).
Before the universe had cooled sufficiently for neutral atoms to
persist, it consisted of a hot plasma of photons, electrons,
protons, baryons and other light nuclei. The tight coupling between
photons and electrons due to Thompson Scattering leads to
oscillations in the hot plasma. As the universe expands and cools,
electrons and protons combine into atoms making the universe
neutral. The pattern of initial perturbations and expanding
wavefronts is seen in the CMB, and is ultimately imprinted on the
matter distribution and should be seen in the spectrum of galaxy
correlations today. The primary representation of these baryon
acoustic oscillations is a feature at the ``sound horizon'' length
$r_{s}$ which is the distance traveled by the acoustic waves by the
time of plasma recombination (Copeland et al. 2006; Albrecht et al.
2006).

The size of baryon acoustic oscillation peak can be used to as a
``standard cosmological ruler'' to constrain the cosmological
parameters (Blake and Glazebrook 2003;  Seo and Eisenstein 2003;
Dolney et al. 2006 ), which was first successfully found by
detecting of a peak in the correlation function of about 50,000
luminous red galaxies over 3800 $deg^2$ in the SDSS (Eisenstein et
al. 2005). This peak can be denoted by a parameter \emph{A}, which
is independent of cosmological models and for a flat universe can be
expressed as
\begin{equation} \label{eq:Ath}
A(z_{BAO};\textbf{p})=\sqrt{\Omega_{m,0}}E(z_{BAO})^{-1/3}[\frac{1}{z_{BAO}}\int_{0}^{z_{BAO}}\frac{dz}{E(z;\textbf{p})}]^{2/3},
\end{equation}
where $\Omega_{m,0}=1-\Omega_{X,0}$ and $z_{BAO}=0.35$. The
observational value is $A=0.469\pm 0.017$. The $\chi^2_{BAO}$ value
is
\begin{equation}
\label{eq:x2BAO}\chi^2_{BAO}(\textbf{p})=\frac{(A(z_{BAO};\textbf{p})-0.469)^2}{0.017^2}.
\end{equation}

\begin{figure}[t]
\centering
  \includegraphics[angle=0,width=90mm]{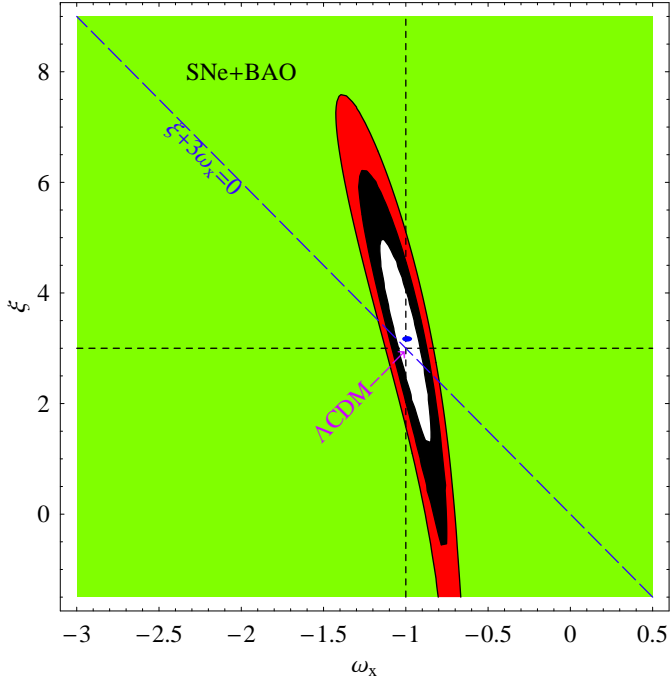}
\caption{
The contours correspond to $68.3\%$, $95.4\%$ and $99.7\%$
confidence levels constrained from joint SNeIa + BAO data analysis
in $(\omega_X, \xi)$ plane.  The results are
$\Omega_{X,0}=0.72\pm0.02$, $\omega_{X}=-0.99\pm0.18$ and
$\xi=3.17\pm1.83$ ($68.3\%$ c.l.).
} \label{fig:BAOCont}
\end{figure}

In this subsection, our analysis consider the SNeIa data combining
with BAO. The best fit values for parameters
$\textbf{p}\equiv(\Omega_{X,0}, \omega_X, \xi)$ can be determined by
minimizing
\begin{equation}
\label{eq:x2tot1} \chi^2_{total}=\chi^2_{SNe}+\chi^2_{BAO}.
\end{equation}
Figure \ref{fig:BAOCont} shows the contours constrained from the BAO
data in addition to the SNeIa data in  $(\omega_X , \xi)$ plane. The
results are $\Omega_{X,0}=0.72\pm0.02$, $\omega_{X}=-0.99\pm0.18$
and $\xi=3.17\pm1.83$ at $68.3\%$ confidence level. Compared to
Figure \ref{fig:SNeCont}, the allowed regions of parameters are not
considerably reduced.

%%%%%%%%%%%%%%%%%%%%%%%%%%%%%%%%%%%%%%%%%%%%%%%%%%%%%%%%%%%%%%%%%%%%%%%%%%%%%%%%%%%%%%%%%%%%%%%%%%%%%%%%%%%%%%%%%%%%%%%%%%%%%%%%%%
%%%%%%%%%%%%%%%%%%%%%%%%%%%%%%%%%%%%%%%%%%%%%%%%%%%%%%%%%%%%%%%%%%%%%%%%%%%%%%%%%%%%%%%%%%%%%%%%%%%%%%%%%%%%%%%%%%%%%%%%%%%%%%%%%%%%

\subsection{Constraints from CMB }

The Cosmic Microwave Background (CMB) observations are playing a key
role in this era of precision cosmology (Barreiro 2009). In 1965
nearly isotropic background of microwave radiation was discovered,
which has provided a wealth of new cosmological data. Subsequently,
a series of experiments, such as COBE (Salopek 1992), Boomerang
(Mauskopf et al. 2000; de Bernardis et al. 2000), MAXIMA (Hanany et
al. 2000), Archeops (Benoit et al. 2003), VSA (Rubino-Martin et al.
2003; Rebolo et al. 2004), DASI ( Kovac et al. 2002; Leitch et al.
2005) and WMAP (Spergel et al. 2003; Spergel et al. 2007; Komatsu et
al. 2009), are designed to detect the CMB. Most notably, the WMAP
satellite has imposed strong constraints on cosmological parameters.

The structure of the anisotropy of the cosmic microwave background
radiation depends on two eras in cosmology, namely last scattering
era $(z_{ls})$ and today $(z=0)$, that can be applied to limit the
model parameters by using the shift parameter $R$. For a flat
universe, $R$ can be expressed as
\begin{equation}
\label{eq:Rth} R(z_{ls};\textbf{p})=\sqrt{\Omega_{m,0}}
\int_{0}^{z_{ls}}\frac{dz}{E(z;\textbf{p})},
\end{equation}
where the last scattering redshift $z_{ls}=1089$. From the 5-year
WMAP data results (Komatsu et al. 2009), one can get the
observational value $R=1.710\pm 0.019$. The $\chi^2_{CMB}$ value is
\begin{equation}
\label{eq:x2R} \chi^2_{CMB}=\frac{(R-1.710)^2}{0.019^2}.
\end{equation}

\begin{figure}[t]
\centering
  \includegraphics[angle=0,width=90mm]{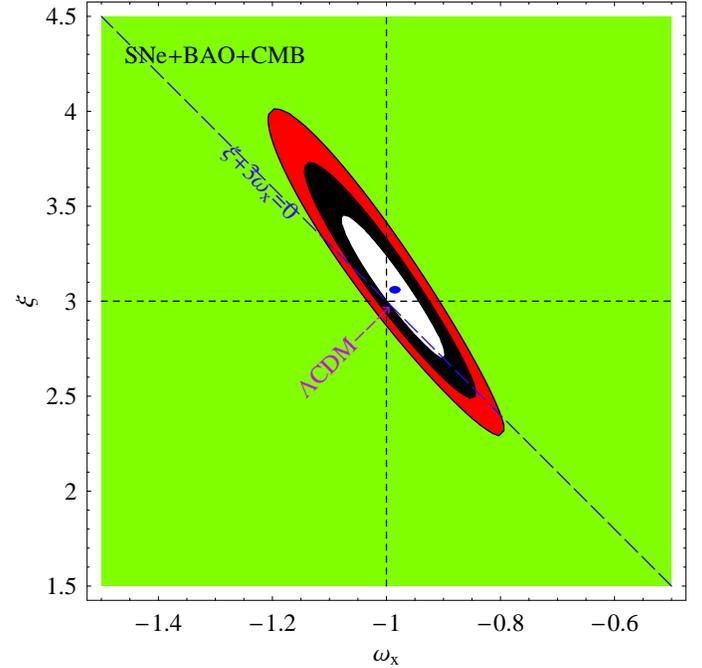}
\caption{
The contours correspond to $68.3\%$, $95.4\%$ and $99.7\%$
confidence levels constrained from joint SNeIa + BAO + CMB data
analysis in $(\omega_X, \xi)$ plane. The values of the parameters
are $\Omega_{X,0}=0.72\pm0.02$, $\omega_{X}=-0.98\pm0.07$ and
$\xi=3.06\pm0.35$ ($68.3\%$ c.l.). It is clear that the $\Lambda$CDM
model is in the $1\sigma$ error region, and the self-similar
solution is out of the $3\sigma$ error region.
} \label{fig:CMBCont}
\end{figure}

We combine the above three data sets to minimize the total
$\chi^2_{total}$,
\begin{equation}
\label{eq:x2tot2}
\chi^2_{total}=\chi^2_{SNe}+\chi^2_{BAO}+\chi^2_{CMB}.
\end{equation}
Figure \ref{fig:CMBCont} shows the contours constrained from the
joint analysis of SNeIa, BAO and CMB data. The values of the
parameters are $\Omega_{X,0}=0.72\pm0.02$, $\omega_{X}=-0.98\pm0.07$
and $\xi=3.06\pm0.35$ ($68.3\%$ confidence level). Compared to
Figure \ref{fig:SNeCont} and Figure \ref{fig:BAOCont}, the allowed
region of $\xi$ is remarkably reduced.

%%%%%%%%%%%%%%%%%%%%%%%%%%%%%%%%%%%%%%%%%%%%%%%%%%%%%%%%%%%%%%%%%%%%%%%%%%%%%%%%%%%%%%%%%%%%%%%%%%%%%%%%%%%%%
%%%%%%%%%%%%%%%%%%%%%%%%%%%%%%%%%%%%%%%%%%%%%%%%%%%%%%%%%%%%%%%%%%%%%%%%%%%%%%%%%%%%%%%%%%%%%%%%%%%%%%%%%%%%%

\begin{figure*}[t]
\centering
  \includegraphics[angle=0,width=180mm]{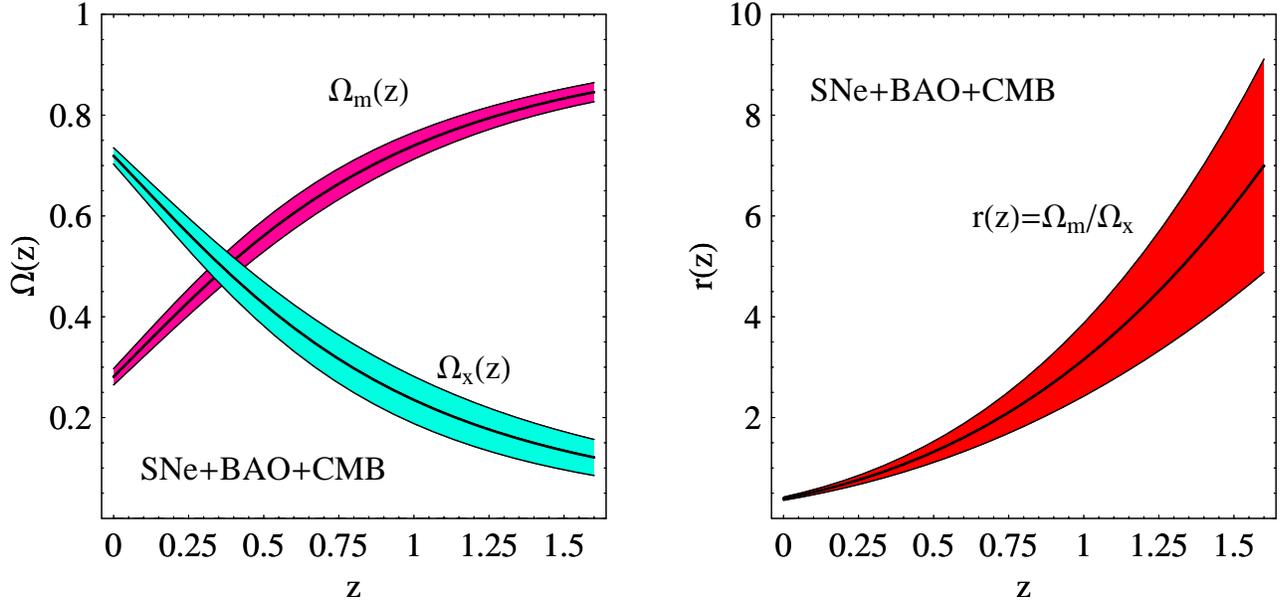}
\caption{
Constrains from the joint analysis of the SNeIa, BAO and CMB data.
The left panel displays the evolutions of $\Omega_m(z)$ and
$\Omega_X(z)$. Wherein, the center solid lines are plotted with the
best fit values, and the shadows denote the $1\sigma$ regions. The
right panel shows the evolution of the ratio of the densities
$r(z)=\Omega_m(z)/\Omega_X(z)$.
} \label{fig:Omega}
\end{figure*}

In addition, Figure \ref{fig:Omega} presents other results
constrained from the joint analysis of SNeIa, BAO and CMB data. The
left panel of Figure 4 displays the evolutions of $\Omega_m(z)$ and
$\Omega_X(z)$. The right panel of Figure 4 presents the evolution of
their ratio $r(z)=\Omega_m(z)/\Omega_X(z)$. It shows that the values
of $\Omega_m(z)$ and $\Omega_X(z)$ are the same order of magnitude
in the redshift range $0.0\leq z \leq 1.6$.

%%%%%%%%%%%%%%%%%%%%%%%%%%%%%%%%%%%%%%%%%%%%%%%%%%%%%%%%%%%%%%%%%%%%%%%%%%%%%%%%%%%%%%%%%%%%%%%%%%%%%%%%%%%%%
%%%%%%%%%%%%%%%%%%%%%%%%%%%%%%%%%%%%%%%%%%%%%%%%%%%%%%%%%%%%%%%%%%%%%%%%%%%%%%%%%%%%%%%%%%%%%%%%%%%%%%%%%%%%%%

\section{Constraints from the transition redshift $z_{T}$ }
The transition redshift (deceleration/acceleration) has been proved
to provide an efficient way for constraining the models (Zhu and
Fujimoto 2004 ,Zhu and Alcaniz 2005). According to the definitions
of the decelerating parameter $q\equiv -(\ddot{a}a)/{\dot{a}}^2$ and
the Hubble parameter $H\equiv \dot{a}/a$, one obtains
\begin{equation}
\label{eq:q1} q=(-\frac{\ddot{a}}{a})/H^2=\frac{d H^{-1}}{d t}-1.
\end{equation}
By using the relations $a_0/a=1+z$ and $E(z)=H/H_0$,
Eq.(\ref{eq:q1}) can be written as
\begin{equation}
\label{eq:q2} q(z)=\frac{1}{2E^2(z)}\frac{dE^2(z)}{dz}(1+z)-1,
\end{equation}
where $E^2(z)$ is given by Eq.(\ref{eq:e2}). The transition redshift
$z_T$, at which the expansion underwent the transition from
deceleration to acceleration, is obtained by solving the equation
\begin{equation}
\label{eq:q3} q(z=z_T)=0.
\end{equation}
From Eq.(\ref{eq:e2}), Eq.(\ref{eq:q2}) and Eq.(\ref{eq:q3}), we
find
\begin{equation}
\label{eq:zt}(1+z_T)^{-\xi}=\frac{\Omega_{X,0}-1}{\Omega_{X,0}(1+3\omega_X)}.
\end{equation}

\begin{figure}[t]
\centering
  \includegraphics[angle=0,width=90mm]{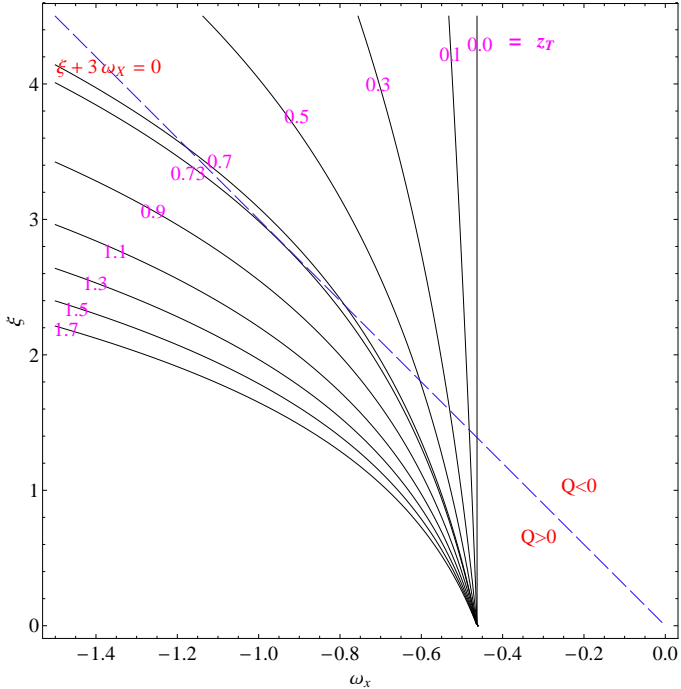}
\caption{
Contour plots of the function $z_T(\omega_X, \xi)$ with a prior
$\Omega_{X,0}=0.72$. The line $\xi+3\omega_X=0$ corresponds to the
standard cosmology without interaction between dark energy and dark
matter. The contour and the line $\xi+3\omega_X=0$ has only one
intersection, corresponding to $(\omega_X, \xi)=(-0.94,2.83)$, with
$z_T=0.73$. Obviously, the contours have no intersections with the
line $\xi+3\omega_X=0$ for $z_T>0.73$.
} \label{fig:TRConstr}
\end{figure}

By considering the results of Section 3, we take a prior on
$\Omega_{X,0}$, i.e., $\Omega_{X,0}=0.72$. In Figure
\ref{fig:TRConstr}, we display the constraints from the transition
redshift $z_T$ in $(\omega_X, \xi)$ plane with the above prior on
$\Omega_{X,0}$. The contour and the line $\xi+3\omega_X=0$ has only
one intersection, corresponding to $(\omega_X, \xi)=(-0.94,2.83)$,
with $z_T=0.73$. Obviously, the contours have no intersections with
the line $\xi+3\omega_X=0$ for $z_T>0.73$. Therefore, if our
universe really turns from deceleration to acceleration at the
redshit $z_T>0.73$, in the framework of this model, we can conclude
that the interaction between dark energy and dark matter should be
taken into account, and the energy is transferred from dark energy
to dark matter. Otherwise, if the transition happens at the redshift
$z_T\leq 0.73$, we just can not ensure whether the interaction is
necessary by using the transition redshift only. In Amendola 2003
and Amendola et al. 2006, they conclude that the acceleration could
have started at high redshift, even up to $z\approx 3$, if dark
energy interacts strongly with dark matter, in contrast, the
standard noninteracting models even hardly reaches $z_T \simeq 1$.
This is consistent with our results. If the transition redshift can
be estimated by a model-independent measurement, it could be used as
a important test to distinguish between coupled and uncoupled
quintessence classes of models.

%%%%%%%%%%%%%%%%%%%%%%%%%%%%%%%%%%%%%%%%%%%%%%%%%%%%%%%%%%%%%%%%%%%%%%%%%%%%%%%%%%%%%%%%%%%%%%%%%%%%%%%%%%%
%%%%%%%%%%%%%%%%%%%%%%%%%%%%%%%%%%%%%%%%%%%%%%%%%%%%%%%%%%%%%%%%%%%%%%%%%%%%%%%%%%%%%%%%%%%%%%%%%%%%%%%%%%%%%

\section{Discussions and Conclusions}
A phenomenological scaling solution $\rho_X\propto \rho_m a^{\xi}$
with minimal underlying theoretical assumptions appears to be a
quite effective tool for analyzing the relationship between the two
dark components of our universe. In this phenomenological model, the
standard cosmology without interaction between dark energy and dark
matter is characterized by $\xi+3\omega_X=0$, whereas
$\xi+3\omega_X\neq 0$ denotes non-standard cosmologies. The value of
$\xi$ quantifies the severity of the coincidence problem while the
special cases $\xi=3$ and $\xi=0$ correspond, respectively, to
$\Lambda$CDM and the self-similar solution without coincidence
problem. Hence, any solution with a scaling parameter $0<\xi<3$
makes the coincidence problem less severe. We have investigated the
constrains imposed by the recent observations. Using the
Constitution Set (397 SNeIa data) solely, we obtain
$\Omega_{X,0}=0.71\pm0.03$, $\omega_{X}=-1.01\pm0.17$ and
$\xi=3.16\pm1.91$ ($68.3\%$ c.l). When BAO data is taken into
account in addition to the SNeIa data, the results are
$\Omega_{X,0}=0.72\pm0.02$, $\omega_{X}=-0.99\pm0.18$ and
$\xi=3.17\pm1.83$ ($68.3\%$ c.l.). Finally, combining SNeIa (397
data), BAO and CMB data, it comes out that
$\Omega_{X,0}=0.72\pm0.02$, $\omega_{X}=-0.98\pm0.07$ and
$\xi=3.06\pm0.35$ ($68.3\%$ c.l.).

Figure \ref{fig:SNeCont}-- \ref{fig:CMBCont} show the observational
contours from SNeIa data, SNeIa + BAO data and SNeIa + BAO + CMB
data in $(\omega_X, \xi)$ plane respectively. From these figures, it
is rather obvious that the SNeIa and BAO data do not provide
stringent constraints on $\xi$, but inclusion of the CMB data
significantly reduces the allowed region of this parameter. This
implies that the high redshift may be able to give tighter
constraint on the parameter $\xi$. In the three cases, they all
display obviously that the self-similar solution $\xi=0$ without
coincidence problem is excluded from the data. As we see from the
three contours, the $\Lambda$CDM model, which corresponds to the
point $(\omega_X, \xi)=(-1, 3)$, is within the $1\sigma$ contour
bound. It shows that the $\Lambda$CDM model still remains a good fit
to the recent observational data, as well as, the coincidence
problem indeed exists and is quite severe. In addition, there is a
tendency in the contours that $\xi$ decreases as $\omega_X$
increases.

The theoretical constrains from the transition redshift $z_T$ show
that if the transition from deceleration to acceleration happens at
the redshift $z_T>0.73$, in the framework of this model, the
interaction between dark energy and dark matter should be taken into
account. On the other hand, if it happens at the redshift $z_T\leq
0.73$, we just can not confirm whether the interaction is necessary
by using the transition redshift only.

Two problems deserve to be pointed out here. Firstly, the line
$\xi+3\omega_X=0$, corresponding to the standard cosmology without
interaction between the two dark components, runs through the
$1\sigma$, $2\sigma$ and $3\sigma$ regions in the $(\omega_X, \xi)$
plane. It denotes that the recent observational data are
insufficient to discriminate between the standard cosmology and the
non-standard cosmology. This problem may be resoluble by using the
transition redshift test, if the transition redshift can be obtain
by a model-independent measurement. Secondly, the data also can not
discriminate between $\omega_X>-1$ and $\omega_X<-1$. In order to
break the degeneracy, we pin our hope on the future observational
data of high redshift SNeIa data from SNAP etc (Albrecht et al.
2006), and more precise CMB data from the ESA Planck satellite
(Balbi 2007), as well as other complementary data, such as Gamma Ray
Bursts data (Schaefer 2007; Liang et al. 2008; Liang and Zhang
2008), the data of X-ray gas mass fraction in cluster (Allen et al.
2004; LaRoque et al. 2006; Allen et al.2008;  Ettori et al. 2009 )
and gravitational lensing data (see Albrecht et al. 2006 and
corresponding references therein).

%%%%%%%%%%%%%%%%%%%%%%%%%%%%%%%%%%%%%%%%%%%%%%%%%%%%%%%%%%%%%%%%%%%
%%%%%%%%%%%%%%%%%%%%%%%%%%%%%%%%%%%%%%%%%%%%%%%%%%%%%%%%%%%%%%%%%%%
%%%%%%%%%%%%%%%%%%%%%%%%%%%%%%%%%%%%%%%%%%%%%%%%%%%%%%%%%%%%%%%%%%%
\acknowledgments
This work was supported by the National Science Foundation of China
 under the Distinguished Young Scholar Grant 10825313 and the Key Project
  Grants 10533010 and 10935013,
 the Ministry of Science and Technology national basic science Program
  (Project 973) under grant Nos. 2007CB815401 and 2010CB833004,
 and the Natural Science Foundation Project of CQ CSTC under grant
  No. 2009BA4050.
JSA acknowledges financial support from CNPq-Brazil under grant
 no. 304569/2007-0 and 481784/2008-0.

\end{document}